\documentclass[useAMS, usenatbib]{mn2e}
\usepackage{graphicx}
\title[Radiation intensity and polarization in atmosphere with chaotic magnetic field] {Radiation intensity and polarization in atmosphere with chaotic magnetic field}
\author[ N. A. Silant'ev, G. A. Alekseeva, Yu. K. Ananjevskaja]{N. A. Silant'ev \thanks {E-mail: nsilant@bk.ru} , G. A. Alekseeva, Yu. K. Ananjevskaja\\
Central Astronomical Observatory at Pulkovo, 196140, Saint-Petersburg, Pulkovskoe shosse 65, Russia}

\begin{document}

\date{ Accepted  ????. Received ????;  in original form ????}
\pagerange { \pageref{firstpage}-- \pageref{lastpage}} \pubyear {2018}
\maketitle
\label {firstpage}

\begin{abstract}
We consider the influence of Faraday rotation in chaotic magnetic field on the intensity and linear polarization of multiple scattered radiation.
 All fluctuations are assumed as Gaussian type and  isotropic. The values of magnetic field are less $10^5$ Gauss. In this case the
parameter $ \omega_B/\omega=(eB/m_ec)/\omega\simeq 0.93\cdot 10^{-8}\lambda$($\mu$m)$B($G$)$ is small and the radiation scatterings are the usual scatterings on free electrons. Only the Faraday rotation influences on the intensity and polarization of radiation. We consider the Milne problem in the electron magnetized atmosphere without the mean magnetic field. We found that chaotic magnetic field does not increase the number of H-functions, describing the Milne problem. There are four H-functions as in the Milne problem for non-magnetized case. The calculations demonstrate that the polarization of outgoing radiation diminishes strongly with the increasing of the level of magnetic fluctuations. The calculations can be used for estimations  of inclination angle $i$ of accretion discs and the level of magnetic fluctuations.
\end{abstract}

$^*$E-mail: nsilant@bk.ru

\section{Introduction}

 The turbulent motions in stellar magnetized atmospheres, accretion discs and other objects are widespread phenomena (see, for example, Rotshtein \& Lovelace 2008 and  Penna et al. 2010).  It is known that in electron plasma the frozen magnetic field occurs (see Landau \&  Lifshitz 1984). There are many MHD - waves in  magnetized plasma, which also acquire  stochastic character (see Priest 1982). The Milne problem corresponds to multiple scattering of non-polarized thermal radiation going from the optically thick level of atmosphere. The polarization occurs  by  the last scatterings before escape the atmosphere.

The Milne problem for polarized radiation in non-magnetized electron atmosphere firstly was solved by Chandrasekhar 1947 and Sobolev 1949 (see also Chandrasekhar 1960 and Sobolev 1969). For regular magnetic field perpendicular to the surface the Milne problem was derived by Silant'ev 1994; Agol \& Blaes 1996; Agol et al. 1998; Shternin et al. 2003. The considered problems are axially symmetric.

The general theory of radiative transfer in stochastic magnetized atmosphere  was given by Silant'ev 2005. More detailed the problem of the derivation of radiative transfer in stochastic media is presented in Silant'ev et al.2017b. The derivation of the system of nonlinear
equations for H-functions in stochastic magnetic fields on the base of the invariance principle was given in Silant'ev 2007.

Here we derived the system of 4  H-functions following to generalized Sobolev's method (see Sobolev 1969; Silant'ev et al. 2017a).
 Note that the Sobolev technique (the resolvent method) also works for the general case of given sources, which does not be calculated by the principle of invariance method. We present detailed data for angular distribution and degree of polarization for many values of magnetic field fluctuations. They can be used to estimate the inclination angle of accretion discs and the level of magnetic fluctuations.

We observe the intensity  $I(\mu,\tau)$ and polarization $Q(\mu,\tau)$ in stochastic atmosphere as  mean values over the time of observation and over the area of observation. Here $\mu={\bf n\cdot N}$ is the cosine of the angle between line of sight ${\bf n}$ and the normal ${\bf N}$ to the surface, $\tau$ is the optical depth in an atmosphere directed  along the normal {\bf N}. These observed values correspond to the average over statistical ensemble of realizations. The stochastic values are assumed to be of  Gaussian type. We denote the averaged values by brackets , for example: $\langle I(\mu,\tau)\rangle\equiv I_0(\mu,\tau)$, $\langle Q(\mu,\tau)\rangle \equiv Q_0(\mu,\tau)$ and  $\langle U(\mu,\tau)\rangle \equiv U_0(\mu,\tau)$.  The fluctuations we denote by primes: $I'(\mu,\tau)$,  $Q'(\mu,\tau)$ and  $U'(\mu,\tau)$.  The averaged fluctuations are equal to zero, $\langle I'\rangle=0$, $\langle Q'\rangle=0$ and $\langle U'\rangle=0$.

Our aim is to derive the  radiative transfer equations for $I_0(\mu,\tau)$,  $Q_0(\mu,\tau)$ and  $U_0(\mu,\tau)$ from exact transfer equations, where
all values are assumed to be stochastic -  the absorption factor $\alpha= \alpha_0 +\alpha'$ and magnetic field ${\bf B}={\bf B_0}+{\bf B'}$, etc. The procedure of derivation is presented in Silant'ev et al. 2017b in detail. According to this method, we average the exponential solutions of the transfer equations without the integral terms. This gives rise to effective absorption factors  $\alpha^{(I)}_{eff}$ for the mean intensity $I_0(\mu,\tau)$ and the effective absorption factor $\alpha^{(pol)}_{eff}$ for the Stokes parameters $Q_0(\mu,\tau)$  and $U_0(\mu,\tau)$. Using the Gaussian type of the ensemble of fluctuations, we obtain $\langle \exp{[-(\alpha_0+\alpha')z]}\rangle=\exp{(-\alpha_{eff}z)}$. Below we give the results of such procedure.

Note that the every turbulence is characterized by the mean length of correlations $R_1$  (the mean radius of curls in the fluid turbulence) and by the dependence of correlation on the distance $R$ between two considered points, characterized by the function $A(R)$. Our theory depends on the integrals from the function $A(R)$. These  integrals we denote as $f_{\alpha}$, $f_{B}$ etc. They depend weakly on the particular forms of function $A(R)$ and are close to  unit. In particular, for $A(R)\sim \exp{(-R/R_1)}$ the value $f_{\alpha}=1$.

Following to the technique of Silant'ev et al. 2017b, we obtain:
\[
\alpha^{(I)}_{eff}=\alpha_0\left(1-\frac{\langle \alpha'^2\rangle}{\alpha_0^2}\,\tau_1f_{\alpha}\right),
\]
\[
\alpha^{(pol)}_{eff}=\alpha^{(I)}_{eff}\,(1+h),
\]
\[
h=\frac{\delta^2\alpha_0\,\tau_1f_B\langle B'^2\rangle}{3\,\alpha^{(I)}_{eff}}\simeq
\]
\begin{equation}
0.213\lambda^4(\mu m)\langle B'^2(G)\rangle \tau_1\frac{\alpha_0}{\alpha^{(I)}_{eff}}f_B.
\label{1}
\end{equation}
\noindent Here $\tau_1=\alpha_0 R_1$, where $\alpha_0=N_e^{(0)}\sigma_T$. The value $N_e^{(0)}$ is the mean number density of free electrons and $\sigma_T=(8\pi/3) r_e^2\simeq 6.5\cdot 10^{-25}$ cm$^2$ is the Thomson scattering cross-section, $r_e=e^2/m_ec^2\simeq 2.82\cdot 10^{-13}$ cm is the classical radius of electron, e is the electron charge. Note that the number density $N_e=N_e^{(0)}+N'_e$. Parameter $\delta$ describes the Faraday rotation of the wave electric field. The angle of rotation $\chi$ is equal to:
\[
 \chi=\frac{1}{2}\delta B_{||}\tau_T,
\]
\begin{equation}
  \delta B_{||}=\frac{3\lambda}{4\pi r_e}\cdot \frac{eB_{||}}{m_e\,c\,\omega}\simeq 0.8\lambda^2(\mu m)B_{||}(G),
\label{2}
\end{equation}
\noindent where wavelength $\lambda$ is taken in microns and the magnetic field in gauss. $ B_{||}={\bf B\cdot n}$ is part of magnetic field along the direction of light propagation ${\bf n}$, $\tau_T=N_e\sigma_Tz $ is the Thomson optical length of $z$. We assume $\langle B'^2_{||}\rangle=\langle B'^2\rangle /3$. 

Eq.(1) shows that absorption factor $\alpha^{(I)}_{eff}$ is less than the mean absorption factor $\alpha_0$, i.e.  the atmosphere with chaotic motions is more transparent than non-turbulent one. In opposite, the absorption factor of polarization parameters $Q$ and $U$ due to the chaotic Faraday rotations became greater than those in non-magnetized medium.  In the Milne problem, where the semi-infinite atmosphere is considered, the first effect is not displayed, but the second is showed strongly.

\section{Radiative transfer equation }

 According to Dolginov et al. 1995 we have the transfer equations:
\[
({\bf n}\nabla)I({\bf n}, {\bf r})=
\]
\[
-\alpha I({\bf n},{\bf r})+N_e\sigma_T B_I({\bf n},{\bf r}),
\]

\[
({\bf n}\nabla)Q({\bf n}, {\bf r})=
\]
\[
-\alpha Q({\bf n},{\bf r})-N_e\sigma_T\delta\,{\bf B\cdot n}\, U({\bf n},{\bf r})+N_e\sigma_T B_Q({\bf n},{\bf r}),
\]
\[
({\bf n}\nabla)U({\bf n}, {\bf r})=
\]
\begin{equation}
-\alpha U({\bf n},{\bf r})+N_e\sigma_T\delta\,{\bf B\cdot n}\, Q({\bf n},{\bf r})+N_e\sigma_T B_U({\bf n},{\bf r}).
\label{3}
\end{equation}
\noindent Here terms $B_I,B_Q$ and $B_U$ describe the scattering on  free electrons. The particular form of these terms is given in
Chandrasekhar 1960.   Considering the Milne problem, we do not write the source terms in this system of transfer equations.

To obtain the system of equations for averaged values $I({\bf n},{\bf r})$, $Q({\bf n},{\bf r})$ and  $U({\bf n},{\bf r})$, we take the value $R({\bf n},{\bf r})=-Q({\bf n},{\bf r})+iU({\bf n},{\bf r})$:
\[
({\bf n}\nabla)R({\bf n}, {\bf r})=
\]
\begin{equation}
-[\alpha + iN_e\sigma_T\delta \,{\bf B\cdot n}]\, R({\bf n},{\bf r})+N_e\sigma_T B_R({\bf n},{\bf r}).
\label{4}
\end{equation}
\noindent Note that the term $({\bf n}\nabla)R({\bf n}, {\bf r})\equiv \mu dR/ds$ characterize the variations along the line of sight ${\bf n}$. It is clear, that
the characteristic length along the line of sight is the free scattering length $s\simeq 1/N_e\sigma_T$. We consider the statistical ensemble of fluctuations on this length. The solution of Eq.(4) without term  $ B_R({\bf n},{\bf r})$ has the form:
\[
R({\bf n}, s)=
\]
\begin{equation}
R(0)\exp{\left[-\int^s_0ds'\,(\alpha(s')+i\,N_e(s')\sigma_T\delta\,{\bf B}(s')\cdot{\bf n})\right]}.
\label{5}
\end{equation}
\noindent This exponential function can be simply averaged according to the technique described in Silant'ev et al. 2017b. To obtain the total averaged equations for $I_0({\bf n}, {\bf r}), Q_0({\bf n}, {\bf r})$ and $U_0({\bf n}, {\bf r})$ we  derived the averages of $\langle N_e\sigma _TB_ I({\bf n},{\bf r})\rangle =\alpha_{eff}B^{(0)}_I$ and $ \langle N_e\sigma_T R({\bf n},{\bf r})\rangle =\alpha_{eff}(1+h)R_0$.

As a result, the system (3) transforms to the system:
\[
({\bf n}\nabla)I_0({\bf n}, {\bf r})=
\]
\[
-\alpha^{(I)}_{eff} I_0({\bf n},{\bf r})+\alpha^{(I)}_{eff} B_{I_0}({\bf n},{\bf r}),
\]
\[
({\bf n}\nabla)Q_0({\bf n}, {\bf r})=
\]
\[
-\alpha^{(I)}_{eff}(1+h) Q_0({\bf n},{\bf r})+ \alpha^{(I)}_{eff} B_{Q_0}({\bf n},{\bf r}),
\]
\[
({\bf n}\nabla)U_0({\bf n}, {\bf r})=
\]
\begin{equation}
-\alpha^{(I)}_{eff}(1+h) U({\bf n},{\bf r})+ \alpha^{(I)}_{eff} B_{U_0}({\bf n},{\bf r}).
\label{6}
\end{equation}
\noindent Note that  every  integral term $B_I({\bf n},{\bf r})$,  $B_Q({\bf n},{\bf r})$ and  $B_U({\bf n},{\bf r})$  depends on all parameters $I,Q,U$.

 Below we consider axially symmetrical problem with ${\bf B}_0= 0$ (the Milne problem). In this case $U({\bf n},{\bf r})\equiv 0$.
 The system of equations for $I_0(\mu,\tau)$ and $Q_0(\mu,\tau)$ has the form:
\[
\mu \frac{d}{d\tau}\left (\begin {array}{c} I_0(\tau,\mu) \\ Q_0(\tau,\mu) \end{array}\right )\equiv \mu \frac{d}{d\tau}{\bf I}_0(\mu,\tau)=
\]
\[
-\left (\begin {array}{c} I_0(\tau,\mu) \\ (1+h)Q_0(\tau,\mu) \end{array}\right )+
\]
\begin{equation}
\frac{1}{2}\hat A(\mu)\int_{-1}^1 d\mu' \, \hat A^T(\mu')\,{\bf I}_0(\mu',\tau).
\label{7}
\end{equation}
\noindent Here we  use  the vector (column) notation  for $I_0(\mu,\tau)$ and $Q_0(\mu,\tau)$. The optical depth is equal to $d\tau=\alpha^{(I)}_{eff}dz$.  The superscript T stands for the matrix transpose. We use the factorization with the matrix (see Silant'ev et al.2017a):

\[
\hat A(\mu)= \hat A_1(\mu)+\hat A_2(\mu)=
\]
\begin{equation}
\left (\begin{array}{rr}1 ,\,\, \,\sqrt{C}(1-3\mu^2) \\ 0 ,\,\,\,3\sqrt{C} (1-\mu^2) \end{array}\right)\equiv
 \left (\begin{array}{rr}1 \, ,\,a(\mu) \\ 0 ,\,\,\,b(\mu) \end{array}\right).
\label{8}
\end{equation}
\noindent Here $C=1/\sqrt{8}=0.353553$,\, $a(\mu)=\sqrt{C}(1-3\mu^2),\,  b(\mu)=3\sqrt{C}(1-\mu^2)$.

The matrices $\hat A_1(\mu)$ and $\hat A_2(\mu)$ have the forms:

\begin{equation}
\hat A_1(\mu)= \left (\begin{array}{rr}1\, , \,a(\mu) \\ 0,\quad 0\,\,\, \end{array}\right),\quad
\hat A_2(\mu)= \left (\begin{array}{rr}0\,\quad , \,\quad 0 \\ 0\,\,, \,\,\,b(\mu) \end{array}\right).
\label{9}
\end{equation}

It is useful  to introduce the vector ${\bf K}(\tau)$:
\begin{equation}
{\bf K}(\tau)\equiv\left (\begin {array}{c} K_1(\tau) \\ K_0(\tau) \end{array}\right )=\frac{1}{2}\int_{-1}^1\,d\mu\,\hat A^T(\mu)\,{\bf I}_0(\mu,\tau).
\label{10}
\end{equation}
\noindent  The matrix product of matrices $\hat A_1$ and $\hat A_2$ is denoted as $\hat A_1\hat A_2$.
 It is easy to verify that Eq.(7) gives rise to  the  conservation law of radiative flux.

\section{ The integral equation for ${\bf K}(\tau)$ }

Formal solution of Eq.(7) gives the integral dependence of  ${\bf I}_0(\mu,\tau)$ on  ${\bf K}(\tau)$.
 Substitution of this solution in Eq.(10) gives homogeneous integral equation for ${\bf K}(\tau)$ :
\begin{equation}
{\bf K}(\tau)|_{hom}=\int_0^{\infty} d\tau'\hat L(|\tau-\tau'|)\,{\bf K}(\tau')|_{hom}.
\label{11}
\end{equation}
\noindent We are  interested in the non-zero solution of this equation. The  kernel $\hat L(|\tau-\tau'|)$  has the form:

\[
\hat L(|\tau-\tau'|)=\int^1_0\frac{d\mu}{\mu}\left[\,\exp{\left(-\frac{|\tau-\tau'|}{\mu}\right)}\hat \Psi_1(\mu)+\right.
\]
\begin{equation}
\left. \exp{\left(-\frac{|\tau-\tau'|(1+h)}{\mu}\right)}\hat \Psi_2(\mu)\right ].
\label{12}
\end{equation}
\noindent The matrices $\hat \Psi_1(\mu)$ and $\hat \Psi_2(\mu)$ have the forms:
\begin{equation}
\hat \Psi_1(\mu)=\frac{1}{2}\hat A_1^T(\mu)\hat A_1(\mu),\quad
\hat \Psi_2(\mu) =\frac{1}{2}\hat A_2^T(\mu)\hat A_2(\mu).
\label{13}
\end{equation}
\noindent Note that $\hat L^T=\hat L$,  $\hat \Psi_1^T=\hat \Psi_1 $ and $\hat \Psi_2^T=\hat \Psi_2$.

The general theory to calculate the vector ${\bf K(\tau)}$ was  presented in Silant'ev et al. 2015.
 Recall, that according to this theory the vector ${\bf K}(\tau)$  has solution through the resolvent matrix  $\hat{R}(\tau,\tau')$:

\begin{equation}
{\bf K}(\tau)=\int_0^{\infty} d\tau'\hat R(\tau,\tau')\,{\bf K}(\tau').
\label{14}
\end{equation}
\noindent Here and in what follows we omit the subscription "hom" . The resolvent matrix obeys the equation:
\begin{equation}
\hat R(\tau,\tau')=\hat L(|\tau-\tau'|)+\int_0^{\infty}d\tau''\hat L(|\tau-\tau''|)\hat R(\tau'',\tau').
\label{15}
\end{equation}
\noindent The kernel $\hat{L}(|\tau-\tau'|)$ of equation for $\hat{R}(\tau,\tau')$ is symmetric: $\hat{L}=\hat{L}^T$ . This gives rise to the relation  $\hat{R}(\tau,\tau')=\hat{R}^T(\tau',\tau)$. The double Laplace transform of $R(\tau,\tau')$ over parameters $a$ and $b$ is equal to:
\begin{equation}
\tilde{\tilde{\hat{R}}}(a,b)=\frac{1}{a+b}[\,\tilde{\hat{R}}(a,0)+\tilde{\hat{R}}^T(b,0)+\tilde{\hat{R}}(a,0)
\tilde{\hat{R}}^T(b,0)].
\label{16}
\end{equation}
\noindent This property demonstrates that the matrix $\hat R(\tau,\tau')$ can be calculated, if we know the matrices  $\hat{R}(\tau,0)$ and
 $\hat{R}(0,\tau)$  (see Silant'ev et al. 2015). For this reason it is useful to study the equation for   $\hat{R}(\tau,0)$, which follows from Eq.(15):
\[
\hat R(\tau,0)=\hat L(\tau)+\int_0^{\infty}d\tau'\hat L(|\tau-\tau'|)\hat R(\tau',0)\equiv
\]
\begin{equation}
\hat L(\tau)+\int_0^{\infty} d\tau'\hat R(\tau,\tau')\hat L(\tau').
\label{17}
\end{equation}
\noindent Using the particular form of $\hat L(|\tau-\tau'|)$ and Eq.(16),  we obtain the relation:

\[
\tilde{\hat R}\left(\frac{1}{a},0\right)=\left(\hat E+\tilde{\hat R}\left(\frac{1}{a},0\right)\right)\int^1_0\frac{d\mu'}{\mu'}\times
\]
\[
\left[\frac{\left(\hat E+\tilde{\hat R}^T\left(\frac{1}{\mu'},0\right)\right)\hat \Psi_1(\mu')}{\frac{1}{a}+\frac{1}{\mu'}}+
 \frac{\left(\hat E+\tilde{\hat R}^T\left(\frac{1+h}{\mu'},0\right)\right)\hat \Psi_2(\mu')}{\frac{1}{a}+\frac{1+h}{\mu'}}\right].
\]
\begin{equation}
\label{18}
\end{equation}
\noindent Here $\hat E$ is the unit matrix. Futher in our theory we use $a=\mu$ and $a=\mu/(1+h)$. Introducing the new values:
\[
\tilde{\hat R}\left(\frac{1}{\mu},0\right)+\hat E\equiv\hat H(\mu)= \left (\begin{array}{rr}A(\mu),\,\,C(\mu) \\ \,D(\mu)\,,\,\,G(\mu) \end{array}\right),
\]
\begin{equation}
\tilde{\hat R}\left(\frac{1+h}{\mu},0\right)+\hat E\equiv\hat F(\mu)= \left (\begin{array}{rr}E(\mu)\, , \,K(\mu) \\ \,M(\mu)\,,\,\,N(\mu) \end{array}\right),
\label{19}
\end{equation}
\noindent we obtain the following equations:
\[
\hat H(\mu)=\hat E+
\]
\begin{equation}
\mu\hat H(\mu)\int^1_0\,d\mu'\left( \frac{\hat H^T(\mu')\hat\Psi_1(\mu')}{\mu'+\mu}+
\frac{\hat F^T(\mu')\hat \Psi_2(\mu')}{\mu'+(1+h)\mu}\right),
\label{20}
\end{equation}
\[
\hat F(\mu)=\hat E+
\]
\begin{equation}
\mu\hat F(\mu)\int^1_0\,d\mu'\left( \frac{\hat H^T(\mu')\hat\Psi_1(\mu')}{(1+h)\mu'+\mu}+
\frac{\hat F^T(\mu')\hat \Psi_2(\mu')}{(1+h)(\mu'+\mu)}\right).
\label{21}
\end{equation}
\noindent Note, that the H-functions $E(\mu)$ and $K(\mu)$ can be obtained from algebraic system of equations, if we know the other functions - $A(\mu)$, $C(\mu)$, $D(\mu)$, $G(\mu)$, $M(\mu)$ and $N(\mu)$. These functions do not occur in the Milne problem.

\section{Formulas for the Milne problem}

The specific feature of the Milne problem is that we are to solve integral  equation (see Eq.(11))  for ${\bf K}(\tau)$ , which has not  the free term.  Using the vector $K(\tau)$, one can obtain the solution of transfer equation (7). The vector ${\bf I}_0(0,\mu)$ describes the emerging radiation. This vector has the form:

\[
{\bf I}_0(0,\mu)=\int_0^{\infty}\frac{d\tau }{\mu}\left[\hat A_1(\mu)\exp{\left (-\frac{\tau}{\mu}\right )}+\right.
\]
\begin{equation}
\left. \hat A_2(\mu)\exp{\left (-\frac{(1+h)\tau}{\mu}\right )}\right]{\bf K}(\tau),
\label{22}
\end{equation}\noindent  i.e. this expression depends on the Laplace transforms of ${\bf K}(\tau)$  over variable $ \tau $.

From Eq.(22) we obtain the particular formulas:
\begin{equation}
I_0(0,\mu)=\frac{1}{\mu}\left[\tilde{K}_1\left(\frac{1}{\mu}\right)+a(\mu) \tilde{K}_0\left(\frac{1}{\mu}\right)\right],
\label{23}
\end{equation}
\begin{equation}
Q_0(0,\mu)=b(\mu)\frac{1}{\mu}\tilde{K}_0\left(\frac{1+h}{\mu}\right).
\label{24}
\end{equation}

 Further we follow to simple approach  of Sobolev 1969, generalizing his method for the  vector
case and magnetized atmosphere.  The final formulas depend on ${\bf K}(0)$. Taking $\tau=0$ in Eq.(11), we obtain:
 \[
{\bf K}(0)=\int_0^{\infty} d\tau\hat L(\tau)\,{\bf K}(\tau)\equiv
\]
\begin{equation}
\int_0^1\frac{d\mu}{\mu}\left[\hat {\Psi}_1(\mu)\tilde {\bf K}\left(\frac{1}{\mu}\right)+\hat {\Psi}_2(\mu)\tilde {\bf K}\left(\frac{1+h}{\mu}\right)\right].
\label{25}
\end{equation}
\noindent Now we find the relations $\tilde {\bf K}\left(\frac{1}{\mu}\right)$ and $\tilde {\bf K}\left(\frac{1+h}{\mu}\right)$ with the matrices  $\hat H(\mu)$ and $\hat F(\mu)$.  Let us get the equation for derivative $d{\bf K}(\tau)/d\tau$, taking into account that the kernel $\hat L(|\tau-\tau'|) $ depends
on the difference $ (\tau-\tau' )$:
\begin{equation}
\frac{d{\bf K}(\tau)}{d\tau}=\hat L(\tau){\bf K}(0)+
\int_0^{\infty} d\tau'\hat L(|\tau-\tau'|)\,\frac{{\bf K}(\tau')}{d\tau'}.
\label{26}
\end{equation}
\noindent  Eq.(26) is non-homogeneous integral equation.  According to the general theory of integral equations (see Tricomi 1957, Smirnov 1964)
 the general solution of Eq.(26) consists of the sum of two terms - the nonzero solution of homogeneous Eq.(11) with some constant $ k $ , i.e. $ k{\bf K}(\tau)$ , and the
solution of non-homogeneous equation (26) with free term $\hat L(\tau){\bf K}(0)$. The latter is proportional to ${\bf K}(0)$ with some factor. This factor  obeys
Eq.(17)  for $\hat R(\tau,0)$, i.e. the general solution of Eq.(26) has the form:

\begin{equation}
\frac{d{\bf K}(\tau)}{d\tau}=k{\bf K}(\tau) +\hat {R}(\tau,0){\bf K}(0).
\label{27}
\end{equation}
Now let us derive the Laplace transforms of this equation. First transform corresponds to  $\exp{(-\tau/\mu)}$, and the second one corresponds to $\exp{(-(1+h)\tau/\mu)}$. Let us consider the first transform in detail. The Laplace transform of the left part of Eq.(27)  is equal to:
\begin{equation}
\int_0^{\infty}d\tau \exp{\left(-\frac{\tau}{\mu}\right)}\frac{d{\bf K}(\tau)}{d\tau}=
-{\bf K}(0)+\frac{1}{\mu}\tilde{\bf K}\left(\frac{1}{\mu}\right).
\label{28}
\end{equation}
\noindent The Laplace transform of the right part of Eq.(27) has the form:
\begin{equation}
k \tilde{{\bf K}}\left(\frac{1}{\mu}\right)+\tilde{\hat R}\left(\frac{1}{\mu},0\right){\bf K}(0).
\label{29}
\end{equation}
\noindent The equality of Eq.(28) with  Eq.(29)  gives rise to the relation:
\begin{equation}
\frac{1}{\mu}\tilde {\bf K}\left(\frac{1}{\mu}\right )=\frac{\hat H(\mu){\bf K}(0)}{1-k\mu}.
\label{30}
\end{equation}
\noindent Here we  used the relation $\tilde{\hat R}(1/\mu,0) =(\hat H(\mu)- \hat E)$ (see Eq.(19) ).

Analogously we obtain the another relation:
\begin{equation}
\frac{1}{\mu}\tilde {\bf K}\left(\frac{1+h}{\mu}\right )=\frac{\hat F(\mu){\bf K}(0)}{1+h-k\mu}.
\label{31}
\end{equation}
 Substituting these formulas into Eqs.(23) and (24), we obtain the final expression for ${\bf I}(0,\mu)$:
\begin{equation}
{\bf I}_0(0,\mu)=\left[\frac{A_1(\mu)\hat H(\mu)}{1-k\mu}+\frac{A_2(\mu)\hat F(\mu)}{1+h-k\mu}\right]{\bf K}(0).
\label{32}
\end{equation}
\noindent The detailed expressions from Eq.(32) are:
\[
I_0(0,\mu)=\beta(\mu)K_1(0)+\gamma(\mu)K_0(0),
\]
\begin{equation}
 Q_0(0,\mu)=b(\mu)[M(\mu)K_1(0)+N(\mu)K_0(0)],
\label{33}
\end{equation}
\noindent where
\[
\beta(\mu)=A(\mu)+a(\mu)D(\mu),
\]
\begin{equation}
 \gamma(\mu)=C(\mu)+a(\mu)G(\mu).
\label{34}
\end{equation}

Thus, the solution of the Milne problem depends on 4 H-functions $\beta(\mu),\gamma(\mu), M(\mu), N(\mu)$ and the components $K_1(0)$
and $K_0(0)$.

 From Eqs.(25), (30) and (31) we obtain the homogeneous algebraic system:
\begin{equation}
{\bf K}(0)=\int_0^1d\mu\left [\frac{\hat \Psi_1(\mu)\hat H(\mu)}{1-k\mu}+\frac{\hat \Psi_2(\mu)\hat F(\mu)}{1+h-k\mu}\right ]{\bf K}(0).
\label{35}
\end{equation}

 This homogeneous equation allows us to obtain only the ratio $K_1(0)/K_2(0)$. So, the expression ${\bf I}(0,\mu)$  contains an arbitrary Const. This Const can be expressed through the observed flux of outgoing radiation. Note that the angular distribution $J(\mu)=I(0,\mu)/I(0,0)$ and the degree of polarization $p(\mu)=Q(0,\mu)/(I(0,\mu)$ are independent of Const.   Note that negative $Q(0,\mu)$ denotes that the wave electric field oscillations are perpendicular to the plane ${\bf (nN)}$.

The necessary condition to obtain ${\bf K}(0)$ is  zero of the determinant of expression (35). We consider the Milne problem in conservative atmosphere. In such case $k=0$.

The system of equations for 4 H-functions $\beta(\mu),\gamma(\mu),M(\mu)$ and $N(\mu)$ is:
\begin{equation}
\beta(\mu)=1+\frac{\mu}{2} \int_0^1d\mu'\,\frac{\beta(\mu)\beta(\mu')+\gamma(\mu)\gamma(\mu')}{\mu+\mu'},
\label{36}
\end{equation}
\[
\gamma(\mu)=a(\mu)+\frac{\mu}{2} \int_0^1d\mu'\,\left( \frac{a(\mu')[\beta(\mu)\beta(\mu')+\gamma(\mu)\gamma(\mu')]}{\mu+\mu'}+\right.
\]
\begin{equation}
\left. \frac{b^2(\mu')[\beta(\mu)M(\mu')+\gamma(\mu)N(\mu')]}{(1+h)\mu+\mu'}\right ),
\label{37}
\end{equation}
\begin{equation}
M(\mu)=\frac{\mu}{2} \int_0^1d\mu'\,\frac{M(\mu)\beta(\mu')+N(\mu)\gamma(\mu')}{\mu+(1+h)\mu'},
\label{38}
\end{equation}
\[
N(\mu)=1+\frac{\mu}{2} \int_0^1d\mu'\,\left(\frac {a(\mu')[(M(\mu)\beta(\mu')+N(\mu)\gamma(\mu')]}{\mu+(1+h)\mu'}+\right.
\]
\begin{equation}
\left. \frac{b^2(\mu')[M(\mu)M(\mu')+N(\mu)N(\mu')]}{(1+h)(\mu+\mu')}\right ).
\label{39}
\end{equation}

These equations can be obtained from Eqs.(20) and (21) by transforms to matrices $\hat A_1 (\mu)\hat H(\mu)$ and $\hat A_2(\mu)\hat F(\mu)$. The Eq.(36) gives rise to the following equation for zeros moments $\overline\beta_0$ and $\overline\gamma_0$:
\[
\overline\beta_0=1+\frac{1}{4}(\overline\beta_0^2+\overline\gamma_0^2),
\]
\begin{equation}
\overline \beta_n=\int_0^1d\mu\,\mu^n\beta(\mu),\quad \overline\gamma_n=\int_0^1d\mu\,\mu^n\gamma(\mu).
\label{40}
\end{equation}
\noindent Considering Eq.(40) as quadratic equation, we obtain $\overline\beta_0=2$, $\overline\gamma_0=0$. For these $\overline\beta_0$ and $\overline\gamma_0$ we find that the determinant of Eq.(35) is really equal to zero. The ratio $K_0(0)/K_1(0)$ is equal to:
\begin{equation}
\frac{K_0(0)}{K_1(0)}=\frac{\int_0^1d\mu\, a(\mu)\beta(\mu)}{1-\int_0^1d\mu \,[\,a(\mu)\gamma(\mu)+b^2(\mu)(M(\mu)+N(\mu))]}.
\label{41}
\end{equation}
\noindent From our calculations  $K_0(0)/K_1(0)=-0.10628$ for $h=0$ and is equal to $ -0.09431 $ for $h=1$. The detailed dependence of  $K_0(0)/K_1(0)$ on parameter $h$ is given in Table 1. The moments $\overline\beta_n, \overline\gamma_n, \overline M_n,\overline N_n$ are given in Table 2 for $h=0$, and in Table 3 for $h=50$.
Note, that we calculated the H-functions $\beta(\mu),\gamma(\mu), M(\mu)$ and $N(\mu)$ by method of successive approximations, using in
the first approximation the values  $\overline\beta_0=2$, $\overline\gamma_0=0$. These values repeat after small numbers of approximations.

\begin{table}
\caption { \small The values  $K_0(0)/K_1(0) $, the angular distribution $J(\mu=1)$ and polarization degree $
p(\mu=0$) for different values of parameter $h$.}
\small
\begin{tabular}{| p{0.6cm} | p{1.8cm}| p{1.3cm}| p{1.5cm}|}
\hline
\noalign{\smallskip}
$ h $ &  $K_0(0)/K_1(0)$ & $J(\mu=1)$ & $ p(\mu=0)\%$  \\
\hline
\noalign{\smallskip}
0        & -0.10628 & 3.063   & 11.713      \\
0.1    & -0.10373 &  3.058   & 10.383      \\
0.2    & -0.10179  & 3.054   & 9.332       \\
0.3    & -0.10023  &  3.051   & 8.478         \\
0.4    & -0.09895  &  3.048   & 7.769         \\
0.5    & -0.09788 &    3.046   & 7.169       \\
0.6    & -0.09697  &  3.044    & 6.656        \\
0.7    & -0.09617 & 3.042    &  6.211         \\
0.8    & -0.09547 &  3.041   & 5.822       \\
0.9    & -0.09486 & 3.040    & 5.479        \\
1       & -0.09431 &  3.039    & 5.174      \\
2       & -0.09088 &  3.033    & 3.320     \\
3       & -0.08917 &   3.029    & 2.441      \\
4       & -0.08813 & 3.028     & 1.930       \\
5       & -0.08742 & 3.026     & 1.595       \\
10     & -0.08577 & 3.023     & 0.853        \\
15     & -0.08512  & 3.022    & 0.582       \\
20     & -0.08478  & 3.022   & 0.441         \\
25     & -0.08456  & 3.021    & 0.356       \\
50     & -0.08410  &  3.021   & 0.180        \\
\hline
\end{tabular}
\end{table}
\normalsize

\begin{table}
\caption { The moments of H-functions $\overline\beta_n,\overline \gamma_n, \overline M_n, \overline N_n$ for $h=0$.}
\small
\begin{tabular}{| p{0.6cm} | p{1.3cm}p{1.3cm}p{1.3cm}p{1.3cm}|}
\hline
\noalign{\smallskip}
$  n    $ & $\overline\beta_n$ & $\overline\gamma_n$ & $\overline M_n $ & $\overline N_n$\\
\hline
\noalign{\smallskip}
0        & 2.00000   & - 0.00000  & 0.08345   & 1.36024  \\
1        & 1.13822   &- 0.12202  & 0.04963    &  0.70915 \\
2        & 0.80123   &- 0.13230  & 0.03505    &  0.47951 \\
3        & 0.61807   & - 0.12497  & 0.02699    &  0.36137 \\
4        & 0.50248   & - 0.11462  & 0.02189     &   0.28935\\
5        & 0.42276   & - 0.10454  & 0.01837     &   0.24085  \\
\hline
\end{tabular}
\end{table}
\normalsize

\begin{table}
\caption { The moments of H-functions $\overline\beta_n,\overline\gamma_n,\overline M_n,\overline N_n$ for $h=50$.}
\small
\begin{tabular}{| p{0.6cm} | p{1.3cm}p{1.3cm}p{1.3cm}p{1.3cm}|}
\hline
\noalign{\smallskip}
$  n    $ & $\overline\beta_n$ & $\overline\gamma_n$ & $\overline M_n $ & $ \overline N_n$\\
\hline
\noalign{\smallskip}
0        & 2.00000   & - 0.00000  & 0.00576   & 1.00606  \\
1        & 1.14062   &- 0.09676   & 0.00362    &  0.49862 \\
2        & 0.80414   &- 0.10262   & 0.00263    &  0.33091  \\
3        & 0.62098   & - 0.09571  & 0.00207   &  0.24700 \\
4        & 0.50526   & - 0.08706  & 0.00169   &  0.19663\\
5        & 0.42536   & - 0.07894  & 0.00143   &  0.16305  \\
\hline
\end{tabular}
\end{table}
\normalsize

\section{Results of calculations and discussion}

The result of calculations of Eqs.(35)-(38) demonstrates that  the linear polarization, emerging from the optically thick magnetized atmocphere, depends very strongly on the parameter $h$ (see Eq.(1)).  We calculated  the angular distribution $J(\mu)=I(\mu)/I(\mu=0)$ and polarization degree $p(\mu)=-Q(\mu)/I(\mu)$. The  peak values  of  the polarization reach at $\mu=0$ and the angular distribution at $\mu=1$. The position angle of emerging radiation corresponds to the wave electric field oscillations parallel to the accretion disc plane, i.e. as in the case of the standard Milne problem.

 Our equations at $h=0$ give rise to  the standard angular distribution and polarization degree  (see Chandrasekhar 1960) with $p_{max}=11,713\%$ and $J_{max}=3.063$. At $h=0.5$ these values are $7.169\%$ and $3.046$, correspondingly (see Table 1). With the increasing of $h$ the polarization tends to zero and the angular distribution tends to standard value  in the Milne problem for the intensity without  taking into account of polarization ( $J_{max}(1)=3.021$).  From Tables 2 and 3 we see, that calculations confirm the exact values $\overline\beta_0=2$ and $\overline\gamma_0=0$ for all values of parameter $h$.

 Table 4 gives the  $J(\mu)$ and $p(\mu)$ - values for $h=0, 0.5$ and $1$ .  In Fig.1 we give the detailed dependence of polarization degree  $p(\mu)$ on parameter $\mu$ for many values of $h$.  The angular distribution $J(\mu)$ is given only  for $h=0$  and $20$.  The value $J(1)$ at $h=20$ is equal to 3.021. Thus, it is seen that $J(\mu)$ practically is independent of parameter $h$.

According to Eqs.(1) and (2), we have:
\begin{equation}
h(\lambda_2)=\left(\frac{\lambda_2}{\lambda_1}\right)^4\,h(\lambda_1),
\label{42}
\end{equation}
\noindent i.e. the dependence of the depolarizing factor $h$ on the wavelength $\lambda$ is very strong. This gives  strong decreasing of polarization degree with increasing of parameter $h$ (see Fig.1). It means also that the polarization decreases strongly with the increase of  the wavelength $\lambda$.


\begin{figure*}
\centering
\setlength\fboxsep{0pt}
\setlength\fboxrule{0.25pt}
\fbox{\includegraphics[width=2\columnwidth]{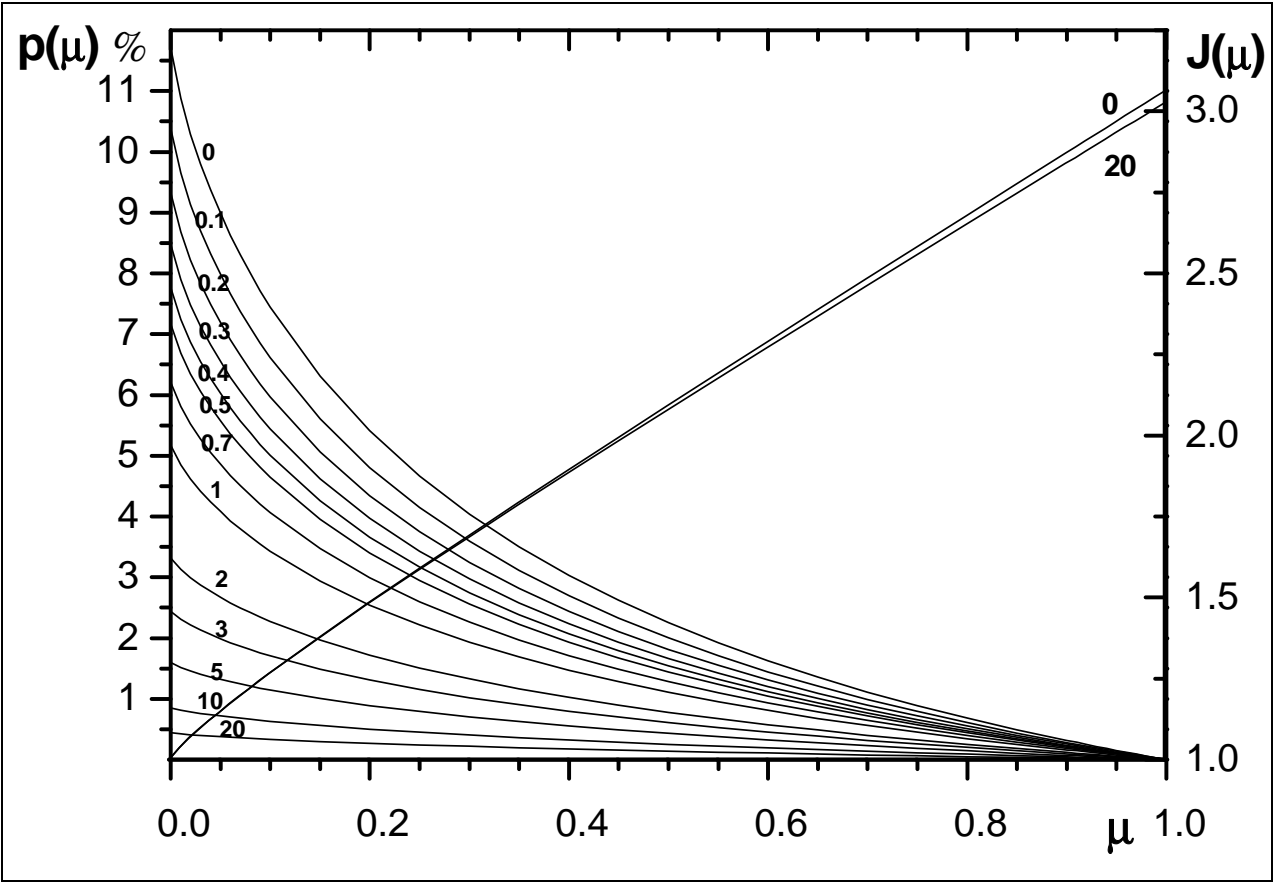}}
\caption{Dependence of polarization $p(\mu)=-Q(\mu)/I(\mu)\%$ on the parameter $h$ (the numbers over curves). The angular distribution $J(\mu)=I(\mu)/I(0)$ is given for $h=0$ and $h=20$.}
\label{fig1.eps}
\end{figure*}

 There are many methods to estimate the inclination angle $i$ ($\cos{i }=\mu $) of an accretion disc plane (see Axon et al. 2008; Marin 2014). All methods assume the different suggestions about physical  picture of considering objects.

 Thus,  Afanasiev et al. (2018) assume that the optically thick accretion discs radiate according to the Milne problem in non-magnetized electron atmosphere. In this case the polarization of continuum
radiation does not depend on the  wavelength.  They considered the objects with such polarization and used the standard Chandrasekhar's dependence of polarization degree on the parameter $\mu$ (see Chandrasekhar 1960).

 Following their method (with the same assumptions) we can consider the accretion discs, where continuum radiation has polarization degree rapidly decreasing with the increase of wavelength $\lambda$. Now the problem is to seek such two parameters - $\mu$ and $h$
in order for obtain the observed polarization degree for many values of wavelength $\lambda$. Below we will show that sometimes this
problem can be solved.

 Table 5 shows that in the atmosphere with chaotic magnetic field  the observed degrees of polarization $p(i)\%$ occur inside various intervals of parameter $h$. Thus, the polarization $p(i)=2\%$ occurs according the Chandrasekhar table only at the inclination $i=57^{\circ}$. But in atmosphere with chaotic magnetic field  this polarization can exist in all values of $h$ from zero up to $h=3$. The existence of these intervals gives possibility to estimate the inclination angle $i$ from two or more observed degrees of polarization.

 As an example, we obtain the inclination angle $i$ for Seyfert 1 nucleus  Mrk 231 (see Smith et al. 2004, Fig. 4). While Smith et al. interpreted the polarization as arising in a jet (polar scattering region), we explore the possibility that  the polarization is due to the Milne problem in optically thick accretion disc with chaotic magnetic field. From Fig.4 in this paper we have: $p(\lambda =0.51 \mu m)=5.5\%,  p(\lambda =0.55  \mu m)=4.8\%, p(\lambda =0.58  \mu m)=4\%$. Using our Fig.1, we obtain $i\simeq 85^{\circ}$. The corresponding values of parameter $h$ are: 0.36, 0.5, 0.62.

Note that  our aim is to demonstrate that the Milne problem in an atmosphere with chaotic magnetic fields can be used in consideration of observed polarization data.

 \section{Conclusion}
The Sobolev's  technique is generalized to describe the Milne problem in the atmosphere with chaotic magnetic fields without the mean value. The radiation equation for this case has different absorption factors for intensity $I$  and the Stokes parameter $Q$.
This case is described by the system of four nonlinear equations for H-functions $\beta(\mu),\gamma(\mu), M(\mu)$ and $N(\mu)$, as it occurs for the Milne problem without  magnetic field. The solution of these equations by the method of successive approximations is very effective, if we use  the exact moments of H-functions $\overline\beta_0=2$ and $\overline\gamma_0=0$. The calculations demonstrate that the chaotic Faraday rotations of the wave electric field  diminishes very effectively the polarization. The results can be used by consideration of linear polarization of radiation in the optically thick chaotically magnetized atmospheres of stars,  accretion discs etc.  Note that in spherical stars with homogeneous atmosphere the total polarization from the star is equal to zero. The total polarization from an star can be observed if the distribution of turbulent magnetic fields is non-homogeneous.
The strong  dependence of polarization on the wave length $\lambda$ may demonstrate that chaotic magnetic field really exists in an atmosphere.

\begin{table}
\caption {\small  The angular distribution $J(\mu)$ and degree of polarization $p(\mu)=-Q(\mu)/I(\mu)$ in \% for $h=0, 0.5 $ and
1 .}
\scriptsize
\begin{tabular}{|p{0.5cm} | p{0.7cm} p{0.7cm}|p{0.7cm} p{0.7cm}| p{0.7cm} p{0.7cm}|}
\hline
\noalign{\smallskip}
$\mu$ & $J_{0}$ & $ p_{0}$ & $J_{0.5}$  &  $ p_{0.5}$ & $J_{1}$ & $p_{1}$     \\
\hline
\noalign{\smallskip}
0       & 1        & 11.713 &     1    & 7.169 &   1     &  5.174 \\
0.01 & 1.036 & 10.878 & 1.037 & 6.692& 1.037 & 4.851    \\
0.02 & 1.066 & 10.295 & 1.067 & 6.351& 1.067 &  4.618   \\
0.03 & 1.094 & 9.805  & 1.094 & 6.061& 1.094 & 4.418     \\
0.04 & 1.120 & 9.374 & 1.121 &  5.805& 1.121 & 4.240  \\
0.05 & 1.146 & 8.986 & 1.146 & 5.573 & 1.146 & 4.079  \\
0.06 & 1.170 & 8.631 & 1.171 & 5.360 & 1.171 & 3.930  \\
0.07 & 1.194 & 8.304 & 1.195 & 5.164 & 1.195 &  3.793   \\
0.08 & 1.218 & 8.000 & 1.218 & 4.980 & 1.218 & 3.664  \\
0.09 & 1.241 & 7.716 & 1.242 & 4.809 & 1.242 & 3.543 \\
0.10 & 1.264 & 7.449 & 1.264 & 4.647 & 1.264 & 3.429  \\
0.15 & 1.375 & 6.312 & 1.375 & 3.956 & 1.375 & 2.939    \\
0.20 & 1.483 & 5.410 & 1.482 & 3.404 & 1.482 & 2.544  \\
0.25 & 1.587 & 4.667 & 1.586 &2.947  & 1.585 & 2.214  \\
0.30 & 1.690 & 4.041 & 1.688 & 2.560 & 1.687 & 1.932  \\
0.35 & 1.791 & 3.502 & 1.789 & 2.225 & 1.787 & 1.688  \\
0.40 & 1.892 & 3.033 & 1.888 & 1.933 & 1.887 & 1.472  \\
0.45 & 1.991 & 2.619 & 1.987 & 1.674 & 1.985 & 1.280 \\
0.50 & 2.091 & 2.252 & 2.085 & 1.443 & 2.083 & 1.107   \\
0.55 & 2.189 & 1.923 & 2.183 & 1.235 & 2.180 & 0.951  \\
0.60 & 2.287 & 1.627 & 2.280 & 1.047 & 2.276 & 0.809  \\
0.65 & 2.385 & 1.358 & 2.376 & 0.876 & 2.373 & 0.679  \\
0.70 & 2.483 & 1.113 & 2.473 & 0.720 & 2.469 & 0.559  \\
0.75 & 2.580 & 0.888 & 2.570 & 0.576 & 2.564 & 0.448  \\
0.80 & 2.677 & 0.682 & 2.665 & 0.443 & 2.660 & 0.346  \\
0.85 & 2.774 & 0.492 & 2.760 & 0.320 & 2.755 & 0.251  \\
0.90 & 2.870 & 0.316 & 2.856 & 0.206 & 2.850 & 0.162  \\
0.91 & 2.890 & 0.282 & 2.875 & 0.184 & 2.869 & 0.144  \\
0.92 & 2.909 & 0.249 & 2.894 & 0.162 & 2.888 & 0.128  \\
0.93 & 2.928 & 0.216 & 2.913 & 0.141 & 2.907 & 0.111  \\
0.94 & 2.947 & 0.184 & 2.932 & 0.120 & 2.925 & 0.094  \\
0.95 & 2.967 & 0.152 & 2.951 & 0.099 & 2.944 & 0.078  \\
0.96 & 2.986 & 0.121 & 2.970 & 0.079 & 2.963 & 0.062  \\
0.97 & 3.005 & 0.090 & 2.989 & 0.059 & 2.982 & 0.046  \\
0.98 & 3.015 & 0.060 & 3.008 & 0.039 & 3.001 & 0.031  \\
0.99 & 3.044 & 0.030 & 3.027 & 0.019 & 3.020 & 0.015  \\
1      & 3.063 & 0        & 3.046 &  0        & 3.039 & 0          \\
\hline
\end{tabular}
\end{table}

\normalsize

\begin{table}
\caption { The dependence of observed degree of polarization $p$\% from values  $h$ and the angle of accretion disc inclination  $i^{\circ}$ .}
\scriptsize
\begin{tabular}{| l | c c c c c c c c c c c c c|}
\hline
\noalign{\smallskip}
$  h   $ & $0$ & $0.5$ & $1.0 $ & $ 2.0$ & $3.0$ & $4.0$ & $5.0$ & $6.0$   \\
\hline
\noalign{\smallskip}
$p=5$\% & $ 77^{\circ}$  & $86^{\circ}$   & $ 90^{\circ}$  & --   & --  & -- & -- &--      \\
4\%         & $ 72^{\circ}$   & $81^{\circ}$  & $87^{\circ}$  &  --  &  -- & -- & --& --              \\
3\%         & $66^{\circ}$   & $76^{\circ}$  & $ 82^{\circ}$  & $ 89^{\circ}$ & --  & -- & --& --      \\
2\%         & $57^{\circ }$  & $67^{\circ}$  & $73^{\circ}$  & $ 82^{\circ}$ & $88^{\circ}$-- &-- &--   \\
1\%  & $45^{\circ}$ & $53^{\circ}$ & $59^{\circ}$ & $ 66^{\circ}$ & $72^{\circ}$ & $78^{\circ}$ &$ 81^{\circ}$ &$85^{\circ}$ \\
0.5\% & $32^{\circ}$ & $40^{\circ}$ & $44^{\circ}$ & $49^{\circ}$ & $57^{\circ}$ & $60^{\circ}$ & $63^{\circ}$ & $67^{\circ}$   \\
\hline
\end{tabular}
\end{table}
\normalsize


\section*{ Acknowledgements.}
This research was supported by the Program of Prezidium of Russian Academy of Sciences N 28 .
 The authors are very grateful to a referee for a number of very useful remarks.

\end{document}